\documentstyle[aps,eqsecnum,epsf,prd,preprint]{revtex}

\begin{document}

\baselineskip=18pt

\preprint{\baselineskip=12pt{\vbox{\hbox{OUTP-98-71P} \hbox{NBI-HE-98-32}
\hbox{hep-th/9811006} \hbox{~~~} \hbox{November 1998}}}}
\title{Induced Dilaton in\\ Topologically Massive Quantum Field
Theory\footnote{\baselineskip=12pt This work was supported in part by the
Particle Physics and Astronomy Research Council (U.K.).}}
\vspace{15mm}
\author{Ian I. Kogan , Arshad Momen}
\address{\vspace{3mm}\baselineskip=12pt Department of Physics -- Theoretical
Physics\\ University of Oxford\\ 1 Keble Road, Oxford OX1 3NP, U.K.\\ {\tt
i.kogan1 , a.momen1 @physics.oxford.ac.uk}}
\author{Richard J. Szabo}
\address{\vspace{3mm}\baselineskip=12pt The Niels Bohr Institute\\ Blegdamsvej
17\\DK-2100 Copenhagen \O, Denmark\\ {\tt szabo@nbi.dk}}
\vspace{1mm}
\maketitle
\begin{abstract}
\baselineskip=14pt

We consider the conformally-invariant coupling of topologically massive gravity
to a dynamical massless scalar field theory on a three-manifold with boundary.
We show that, in the phase of spontaneously broken Lorentz and Weyl symmetries,
this theory induces the target space zero mode of the vertex operator for the
string dilaton field on the boundary of the three-dimensional manifold. By a
further coupling to topologically massive gauge fields in the bulk, we
demonstrate directly from the three-dimensional theory that this dilaton field
transforms in the expected way under duality transformations so as to preserve
the mass gaps in the spectra of the gauge and gravitational sectors of the
quantum field theory. We show that this implies an intimate dynamical
relationship between $T$-duality and $S$-duality transformations of the quantum
string theory. The dilaton in this model couples bulk and worldsheet degrees of
freedom to each other and generates a dynamical string coupling.

\end{abstract}

\vfill
\pagebreak

\font\mathsm=cmmi9
\font\mybb=msbm10 at 12pt
\def\bb#1{\hbox{\mybb#1}}
\font\mybbs=msbm10 at 9pt
\def\bbs#1{\hbox{\mybbs#1}}
\def\e{{\rm e}}

\newcommand{\complex}{{\bb C}} 
\newcommand{\complexs}{{\bbs C}} 
\newcommand{\zed}{{\bb Z}} 
\newcommand{\real}{{\bb R}} 
\newcommand{\reals}{{\bbs R}} 
\newcommand{\zeds}{{\bbs Z}} 
\newcommand{\be}{\begin{equation}}
\newcommand{\ee}{\end{equation}}
\newcommand{\bea}{\begin{eqnarray}}
\newcommand{\eea}{\end{eqnarray}}
\newcommand{\ra}{\rightarrow}
\newcommand{\tr}{\mbox{tr}}
\newcommand{\Tr}{\mbox{Tr}}
\newcommand{\al}{\alpha}
\newcommand{\bt}{\beta}
\newcommand{\bz}{{\bar{z}}}
\newcommand{\del}{\Delta}
\newcommand{\Th}{\Theta}
\newcommand{\td}{\tilde{\del}}
\newcommand{\g}{\gamma}
\newcommand{\bchi}{\bbox{\chi}}
\newcommand{\nn}{\nonumber}

\section{Introduction}

The relationship between gravity in a three-dimensional spacetime and
two-dimensional quantum gravity has been a topic of much interest over the last
decade \cite{verlinde,carkog,iangrav,carlip,kogan,ashw,chvd,kpz}. Of particular
interest is topologically massive gravity \cite{djt} whose relation to
Liouville theory \cite{carlip,kogan,ashw,kpz} extends the general
correspondence between topological gauge theories in three dimensions and
two-dimensional conformal field theories \cite{witten}. It also describes the
gravitational sector of topological membrane theory \cite{iangrav,otm} (see
\cite{ianrev} for a recent review) which reformulates string theory by filling
in the string worldsheet and viewing it as the boundary of a three-manifold.
Many aspects of string dynamics have intriguing interpretations when
represented in terms of the dynamics of gauge and gravitational fields in the
bulk. In this paper we will describe how to incorporate the dilaton field of
string theory into this prescription.

In non-critical string theory, the target space tachyon operator sets the
ultraviolet scale on the string worldsheet and it depends on the dilaton field
$\phi$ and also the Liouville field. However, there is a dilatonic coupling in
the string sigma-model action whose vacuum expectation value is proportional to
the Euler character of the worldsheet and which thereby sets the string
coupling constant $g_s$ according to
\be
g_s=\left\langle\e^\phi\right\rangle
\label{gsdef}\ee
One can change $g_s$ by shifting $\phi$ and thus naively spoil the conformal
invariance of the theory at the quantum level. This also affects the tachyon
operator, and hence the worldsheet scale, so that the dilaton field in this way
controls the scale transformation properties of the string theory. This
property is particularly important for the invariance of the string theory
under duality transformations. If a direction of the $D$ dimensional target
space is compactified on a circle of radius $R_0$, then $T$-duality maps this
circle onto its dual of radius $R_0^*=\alpha'/R_0$, where $\alpha'$ is the
string Regge slope. The expectation value of the dilaton must shift so as to
leave unchanged the corresponding $D-1$ dimensional Planck scale,
\be
\frac{R_0^*}{\kappa_{(D)}^{*\,2}}=\frac{R_0}{\kappa_{(D)}^2}
\label{Pscaleinv}\ee
where $\kappa_{(D)}$ is the $D$ dimensional gravitational coupling. For
instance, for a non-linear sigma-model in a compact target space with metric
$G$ that admits an isometry, the dilaton field is constant along the direction
of the corresponding Killing vector and transforms under $T$-duality as
\cite{buscher}
\be
\phi\to\phi^*=\phi+\frac12\ln\frac{G_{00}}{\alpha'}
\label{buscherlaw}\ee
where $G_{00}=R_0^2$ is the component of the metric tensor in the direction of
the isometry generator.

On the other hand, the natural scale of the bulk gravitational theory is the
three-dimensional Planck mass which sets the coupling constant of topologically
massive gravity \cite{deser}, and the induced worldsheet scale is set by the
topological graviton mass \cite{kogan}. This suggests that within this latter
theory there could exist a three-dimensional version of the string dilaton
field. Namely, the coupling of strings to a dilaton may be related to a
three-dimensional gravitational model with a fluctuating Planck mass. In this
paper we will show precisely how this is accomplished. We will consider the
conformally-invariant coupling of topologically massive gravity to a dynamical,
massless scalar field in the bulk \cite{deserscale}. In the phase of
spontaneously broken Weyl invariance, the vacuum expectation value of the
scalar field induces a mass for the graviton and gives a model reminescent of
old theories of induced gravity \cite{adler}. When the model is written in a
first-order (phase space) formalism for the gravitational fields \cite{xiang},
an induced two-dimensional $SL(2,\real)$ gauged WZNW model emerges, in the
phase of spontaneously broken $SO(2,1)$ gauge symmetry of the three-dimensional
theory, which is well-known to be related to Liouville theory \cite{alex}. In
the present framework, however, we shall find that the extra coupling to the
scalar field induces a scale-dependent deformation of the usual Liouville
theory which shares all of the properties of a (constant) dilaton term in a
string sigma-model. In particular, by further coupling a bulk topologically
massive gauge theory \cite{djt} to the scalar field (representing the inclusion
of a string sigma-model action), we show that the requirement of invariance of
the bulk physical spectrum of the quantum field theory under $T$-duality
transformations of the associated target space leads to the usual form of the
dilaton shift (\ref{buscherlaw}), in much the same way that the string
theoretical requirement (\ref{Pscaleinv}) does. In fact, as we will show, this
correspondence suggests a remarkable equivalence between $T$-duality and
$S$-duality transformations of the quantum string theory.

The origin of the dilaton field in string theory is somewhat mysterious, since
in a sense it is really just a scaffold for producing the string coupling
constant $g_s$. Although in the following we will not reproduce the complete
dilaton vertex in the topological membrane picture, we do manage to capture
that part of it which is essential for the qualitative features of the dilaton
field in string theory. In particular the ensuing construction represents the
proper incorporation of a string coupling into the topological membrane
formulation of string theory, and the dilaton thereby induced by the bulk
theory has a nice dynamical origin in terms of the geometry and the propagating
particles in the three-manifold.

One crucial aspect of the deformed Liouville theory that we obtain is that it
is {\it not} an intrinsically two-dimensional model. The breaking of the
Lorentz and conformal symmetries of the three-dimensional quantum field theory
induces propagating massive gauge and graviton degrees of freedom in the bulk.
The massless scalar field couples bulk and boundary degrees of freedom in such
a way that the three-dimensional dynamics controls the properties of the
induced dilaton field. This immediately leads to the possibility of having a
dilaton field which depends explicitly on the worldsheet coordinates and leads
to a {\it dynamical} string coupling constant (\ref{gsdef}). The induced
dilaton that we find is in fact related to the target space tachyon operator,
so that in this case $g_s$ is a dynamical field in target space that controls
the size of the spacetime. This is one of the basic features of 11-dimensional
$M$-theory \cite{wittenM} whereby the string coupling constant of
ten-dimensional type-IIA superstring theory is related to the radius of the
eleventh dimension. In the present case we also find that the bulk dynamics
induces a sort of new dimension into the model, in addition to the string
embedding fields and the extra dimension induced by the Liouville field. This
gives a potential dynamical origin for the extra dimension of spacetime
inherent in $M$-theory from the basic point of view of fundamental string
fields which could be relevant to the dynamics of the 11-dimensional theory
itself.

The arrangement of this paper is as follows. In section 2, we describe the
model whereby topologically massive gravity is conformally coupled to
a scalar field theory and present the derivation of a ``deformed''
two-dimensional Liouville gravity induced on the boundary by the bulk action,
following \cite{kogan} for the most part. In section 3 we show that with an
appropriate boundary condition on the scalar field one can reproduce the
dilaton vertex operator on the boundary. In section 4 we consider the
additional conformally invariant coupling to topologically massive gauge
theories and derive the transformation laws (\ref{buscherlaw}) directly from
the bulk theory. Finally, in section 5 we discuss the possibility of inducing a
dynamical dilaton field on the worldsheet and its potential relevance to the
dynamics of $M$-theory.

\section{From Conformally-coupled Topologically Massive Gravity to Deformed
Liouville Theory}

In this section, we will consider the action for
 topologically massive gravity conformally coupled to a scalar field theory
 defined on a three-manifold with  boundary.
  Following the derivation of two-dimensional quantum gravity from ordinary
topologically massive gravity \cite{carlip,kogan,ashw}, we will derive
 a deformed Liouville theory induced on the two dimensional boundary
by the bulk theory.

\subsection{Definition of the Bulk Theory}

Consider the action for topologically massive gravity defined on an orientable
three-dimensional Minkowski-signature manifold $\cal M$ without boundary,
\bea
S_{TMG}[e,\omega]=\kappa\int_{\cal M}d^3x~\epsilon^{\mu \nu \lambda}
 e_\mu^a  R_{\nu
\lambda}^a + \frac{k}{8 \pi} \int_{\cal M}d^3x~ \epsilon^{\mu
 \nu \lambda} \left( \omega_\mu^a\,\partial_\nu\omega_\lambda^a +
\mbox{$\frac{2}{3}$}\,\epsilon^{abc} \omega_\mu^a  \omega_\nu ^b
\omega_\lambda^c \right),
\label{TMG}\eea
where $\omega_\mu^a=\epsilon^{abc} \omega_\mu^{bc}$ is the $so(2,1)$ Lie
algebra  valued spin-connection and
\be
R^a = d\omega^a +\epsilon^{abc} \omega^b \wedge \omega^c
\label{curvature}\ee
is its curvature (we will use the differential form
and component notations interchangeably). The first term in (\ref{TMG}) is the
Einstein-Hilbert action written in the first-order formalism, with $\kappa$ the
three-dimensional Planck mass, while the second term is the gravitational
Chern-Simons action. Unlike the first order Palatini action, the spin
connection
in (\ref{TMG}) is not an independent variable but is related to the
triads $e_\mu^a$ and the inverse triads $E^\nu_a$ by the Cartan-Maurer formula
\be
\omega_\mu^a = \epsilon^{abc}\left[ E^{\nu\,b} ( \partial_\mu e_\nu^c -
 \partial_\nu e_\mu^c) - \mbox{$\frac{1}{2}$}\, E^{\rho \,b} E^{\sigma \,c} (
\partial_\rho e_{\sigma}^d - \partial_\sigma e_{\rho}^d ) e_\mu^d \right]
\label{spincon}
\ee
with $e_\mu^a\otimes e_{\nu\,a}=g_{\mu \nu}$ and $E^a_\mu\otimes e^\mu_b=
\delta^a_b$.

Consider a Weyl transformation of the
metric of $\cal M$ (or equivalently a rescaling of the triad fields),
\bea
g_{\mu \nu}(x) \ra \hat{g}_{\mu \nu}(x)  = \Phi(x)^4 g_{\mu
\nu}(x)~~~~~~,~~~~~~
e_\mu^a(x) \ra \hat{e}_\mu^a(x)  = \Phi(x)^2 e_{\mu}^a(x)
\label{scale}
\eea
where $\Phi(x)$ is some scalar field on $\cal M$, so that
\be
\omega^{a}_\mu \ra \hat{\omega}_\mu^{a} = \omega_\mu^{a} +
\epsilon^{abc} E^{\nu  b }e_\mu^{c}\,\partial_\nu  \ln \Phi
\label{spintrans}
\ee
Under the transformations (\ref{scale},\ref{spintrans}), the gravitational
Chern-Simons action is invariant but the Einstein-Hilbert term changes, so that
the total action (\ref{TMG}) transforms as\footnote{\baselineskip=12pt Note
that one could also include a cosmological constant term in the action
(\ref{TMG}).  Under the conformal transformation (\ref{scale})
this term would induce a $\Phi^6$ potential in the action (\ref{CTMG}). This
situation is relevant to the corresponding construction for Einstein gravity on
$AdS_3$ spacetimes which also induces Liouville theory \cite{chvd}. We shall
not discuss this aspect in this paper.}
\bea
S_{TMG}\left [ e,\omega \right]&\ra&S_{TMG}\left[ \hat{e}, \hat{\omega} \right]
\equiv S_{CTMG}[e,\omega;\Phi]\nonumber \\
&=&\kappa\int_{\cal M}d^3x~ \epsilon^{\mu \nu \lambda}\,\Phi^2\,e_\mu^a  R_{\nu
\lambda}^{a}+\frac{k}{8 \pi} \int_{\cal M}d^3x~\epsilon^{\mu \nu \lambda}
\left(\omega_\mu^a\,\partial_\nu\omega_\lambda^{a}+
\mbox{$\frac{2}{3}$}\,\epsilon^{abc} \omega_\mu^a \omega_\nu ^b
\omega_\lambda^c \right)\nonumber\\& &+\,8\kappa\int_{\cal M}d^3x~\sqrt
g~g^{\mu\nu}\,\partial_\mu\Phi\,\partial_\nu\Phi
\label{CTMG}\eea
In the naive vacuum $\langle\Phi^2\rangle=0$, the spectrum of the quantum field
theory (\ref{CTMG}) contains a massless scalar particle but no graviton degrees
of freedom, so that in this phase the model is equivalent to the pure
conformally-invariant gravitational Chern-Simons theory which is a topological
$SO(2,1)$ gauge theory. A more interesting case is when the conformal symmetry
of the pure Chern-Simons action is spontaneously broken and induces the
Einstein-Hilbert term by a sort of Higgs mechanism \cite{deserscale,adler}. If
the scalar field $\Phi(x)^2$ has a non-zero vacuum expectation value, then one
can gauge it away by a Weyl transformation (\ref{scale}) with conformal factor
$\Omega(x)^2=\langle\Phi^2\rangle/\Phi(x)^2$. Then, with the rescaling
$\Phi\to\Omega\,\Phi$, (\ref{CTMG}) becomes the topologically massive gravity
action (\ref{TMG}) for the fields $\hat e,\hat\omega$ with Planck mass
$\hat\kappa=\kappa\,\langle\Phi^2\rangle$. This shows that the Hartree-Fock
average of the
fluctuating field $\Phi$ is related to the topological graviton mass $M_g$ via
\be
M_g=\frac{8\pi\kappa\,\langle\Phi^2\rangle}k
\label{plmass}\ee
In other words, in the theory (\ref{CTMG}) the background field $\Phi$ (or
rather its vacuum expectation value generated by zero-point quantum
fluctuations) sets the mass scale of the bulk theory. Since the perturbation
expansion parameter of topologically massive gravity is the
super-renormalizable coupling constant $M_g/\kappa=8\pi\langle\Phi^2\rangle/k$
\cite{deser}, it follows that $\Phi$ also determines the effective coupling
constant for the model (\ref{CTMG}).\footnote{\baselineskip=12pt Note that we
can absorb the mass scale $\kappa$ into the field $\chi=4\sqrt{\kappa} \,\Phi$
so that $\chi$ has the correct canonical
dimension $\frac{1}{2}$ for a bosonic field in three dimensions.} This
parallels the case in string theory where the vacuum expectation value of the
dilaton field sets the string coupling constant (\ref{gsdef}). In what follows
we will make this correspondence more precise.

The action (\ref{CTMG}) is invariant under the conformal transformations
\bea
\Phi(x)& \ra& \Omega(x)\,\Phi(x) \nonumber \\
g_{\mu \nu}(x)&\ra&\Omega(x)^{-4}\,g_{\mu
\nu}(x)\nn\\e_\mu^a(x)&\ra&\Omega(x)^{-2}\,e^a_\mu(x)\nn\\
\omega_\mu^a(x)&\to&\omega_\mu^a(x)+\epsilon^{abc}E^{\nu b}(x)e_\mu^c(x)\,
\partial_\nu\ln\Omega(x)
\label{confmetric}
\eea
and we shall refer to the model (\ref{CTMG}) as conformally-coupled
topologically massive gravity. However, if $\cal M$ has a non-empty boundary
$\partial\cal M$, then the conformal symmetry of the pure bulk action
(\ref{CTMG}) is explicitly broken. Under the local scale transformation
(\ref{scale}), the Einstein-Hilbert part of the action induces an extra
boundary term,
\be
S_{TMG}[e,\omega] \ra S_{CTMG}[e,\omega;\Phi] -8\kappa\oint_{\partial\cal
M}\Phi\,\partial_\perp\Phi
\label{boundphi}
\ee
where $\partial_\perp$ denotes the normal derivative to the boundary of $\cal
M$. To eliminate this term from the action one would have to impose a Neumann
boundary condition for the field $\Phi$ on $\partial\cal M$. This will be
discussed in section 5. Here we shall impose a Dirichlet boundary condition on
the scalar field,
\be
t^\alpha\,\partial_\alpha \Phi =0~~~~~~{\rm or}~~\Phi(x)\Bigm|_{\partial\cal
M}= {\rm constant}
\label{bc1}
\ee
where $t^\alpha$ is a unit vector along the boundary $\partial\cal
M$. The choice of boundary condition (\ref{bc1}) will ensure that the induced
dilaton field that arises is not an explicit function of the worldsheet
coordinates (but generally only an implicit one through its dependence on the
target space fields and the Liouville field). As a consequence of the Dirichlet
boundary condition, the full conformal symmetry group of the three-dimensional
theory is broken down to the subgroup of conformal transformations
(\ref{confmetric}) which are constant on the boundary of $\cal M$. Thus the
conformal symmetry group of conformally-coupled topologically massive gravity
will induce the group of global scale transformations of the induced
two-dimensional field theory.

\subsection{Derivation of the Induced Boundary Theory}

The action for conformally-coupled topologically massive gravity can be written
in the form
\bea
S_{CTMG}[e,\omega,\Phi,\lambda]&=&\kappa\int_{\cal M} \Phi^2\,
e^a \wedge R^a + \frac{k}{8 \pi} \int_{\cal M}
\left( \omega^a \wedge d \omega^a +
\mbox{$\frac{2}{3}$}\,\epsilon^{abc} \omega^a \wedge \omega^b
\wedge \omega^c \right)+8\kappa\int_{\cal M}(d \Phi)^2 \nonumber \\
& &-\,8\kappa\oint_{\partial\cal M} \Phi\,\partial_\perp\Phi+\int_{\cal
M}\Phi^2\,\lambda^a \wedge \left( de^a +\epsilon^{abc} \omega^b\wedge
e^c+\epsilon^{abc}e^b\wedge\omega^c\right)
\label{2.1}\eea
where the spin-connection $\omega$ and the triad field
$e$ are to be treated as independent variables. Here $\lambda$ is a
Lagrange multiplier field that enforces the torsion-free constraint
(\ref{spincon}) on the geometry and which is invariant under the Weyl
transformations (\ref{confmetric}). The generally-covariant action (\ref{2.1})
is invariant under the restricted conformal transformations (\ref{confmetric})
and also under the local $SO(2,1)$ Lorentz transformations whose infinitesimal
forms are
\bea
\delta_\theta e^a &=& \epsilon^{abc} \theta^b e^c \nn \\
\delta_\theta\lambda^a& =&\epsilon^{abc} \theta^b \lambda^c \nn \\
\delta_\theta\omega^a& =& -\mbox{$\frac{1}{2}$}\left( d \theta^a + 2
\epsilon^{abc} \theta^b\omega^c\right)\nn\\\delta_\theta\Phi&=&0
\label{2.2}\eea
Note that one can rewrite (\ref{2.1}) up to an overall constant as an
$SL(2,\real)$ gauge theory by simply rescaling the action according to the
trace relationship  $\Tr ( T_a T_b) = 4 \,\tr (\tau_a\tau_b)$, where $T_a$ and
$\tau_a$ are the generators of the fundamental representations of $SO(2,1)$ and
$SL(2,\real)$, respectively.

If $\partial{\cal M}\neq\emptyset$, then one needs to augment the
action with appropriate boundary terms to ensure that the resulting path
integral formulation of the quantum field theory has a semi-classical
approximation \cite{carlip,kogan,cooper} (this is equivalent to selecting the
necessary boundary conditions to solve the field equations). After integrating
the Einstein-Hilbert term by parts, the action (\ref{2.1}) can be rewritten as
\be
S_{CTMG}[e,\omega,\Phi,\beta]=S_{\rm
bulk}[e,\omega,\Phi,\beta]+\kappa\oint_{\partial\cal M} \Phi^2~\tr\left(\omega
\wedge e \right) -2\kappa
\oint_{\partial\cal M} \Phi\,\partial_\perp\Phi
\label{2.3.5}\ee
where
\bea
S_{\rm bulk}[e,\omega,\Phi,\beta]&=&\int_{\cal
M}\tr\left[\Phi^2\beta\wedge(de+\omega\wedge e-e\wedge\omega)+\frac
k{8\pi}\left(\omega\wedge
d\omega+\mbox{$\frac23$}\,\omega\wedge\omega\wedge\omega\right)\right]\nn\\&
&+\,\kappa\int_{\cal
M}\tr\Bigl[\Phi\,e\wedge\omega\wedge(2\,d\Phi+\Phi\,\omega)\Bigr]+
2\kappa\int_{\cal M}(d\Phi)^2
\label{bulkaction}\eea
is the contribution from the bulk parts of the fields, and the field
\be
\beta^a=\lambda^a +\kappa\,\omega^a
\label{betadef}\ee
transforms like the spin-connection under both Lorentz and Weyl
transformations. Under variations of the fields which do not necessarily vanish
 on the boundary, the variation of the action (\ref{2.3.5}), restricted to the
boundary, is given by
\be
\delta S_{CTMG}[e,\omega,\Phi,\beta]\Bigm|_{\partial\cal M} =-
\oint_{\partial\cal M} \tr \left[ \Phi^2\beta \wedge \delta e + \frac{k}{8
\pi}\,
\omega  \wedge \delta \omega -\kappa\Phi^2 (\delta \omega \wedge e +
\omega \wedge \delta e )\right].
\label{2.4}\ee
Note that the field $\Phi$ is not varied on the boundary due to the Dirichlet
boundary condition (\ref{boundphi}). Eq. (\ref{2.4}) shows that we need to add
appropriate surface terms to the action to cancel
the boundary variations of the various fields.
The precise form of these terms depends on the boundary conditions imposed on
the various phase space variables.
Choosing a complex structure on $\partial\cal M$, from (\ref{2.4}) it follows
that there are three sets of canonical pairs, namely $(\beta_z, e_\bz ), (
\bt_\bz, e_z)$ and $ (\omega_z , \omega_\bz)$. Thus, we need to specify
boundary
conditions on one of the components from each canonical pair.

Accordingly, let us make the following choice of boundary
conditions on $\partial\cal M$,
\be
\delta e_z=\delta \beta_z=
\delta \omega_\bz =0
\label{2.5}\ee
With these boundary conditions the terms required to
be added to the action are
\bea
S_B[e,\omega,\Phi,\beta]=-\oint_{\partial\cal M}d^2z~\tr
\left[\frac{k}{8\pi}\,\omega_z \omega_\bz -\Phi^2 e_\bz\left( \bt_z
-\kappa\omega_z\right) \right]
\label{2.5.1}\eea
so that the total action reads
\bea
S_{T}[e,\omega,\Phi,\beta]&=&S_{CTMG}[e,\omega,\Phi,\beta] +
S_B[e,\omega,\Phi,\beta]\nn \\
&=&S_{\rm bulk}[e,\omega,\Phi,\beta]-2\kappa\oint_{\partial\cal M}
\Phi\,\partial_\perp\Phi\nn-\oint_{\partial\cal M}d^2z~\tr \left[
\frac{k}{8\pi}\,
\omega_z\omega_\bz-\Phi^2\left(e_\bz\bt_z-\kappa
e_z\omega_\bz\right)\right]\nn\\& &~~~~
\label{2.7}\eea
Under the $SL(2,\real)$ gauge transformations
\bea
e &\ra& g^{-1} e g \nn \\
\beta&\ra& g^{-1}(d+\beta)g \nn \\
\omega &\ra& g^{-1}(d+ \omega )  g\nn\\ \Phi&\to&\Phi
\label{2.8}\eea
the action (\ref{2.7}) is not invariant because the Chern-Simons
term in (\ref{bulkaction}) and the $\omega_z\omega_\bz$ term in (\ref{2.7})
both induce additional boundary terms \cite{witten}. Under the transformations
(\ref{2.8}) a chiral gauged $SL(2,\real)$ WZNW model is induced on the
boundary. The resulting action is given  by
\bea
S_T[e,\omega,\Phi,\beta]&\ra&S_T^-[g;e,\omega,\Phi,\beta]\nn\\&
&=S_T[e,\omega,\Phi,\beta]+ \frac{k}{8 \pi}\oint_{\partial\cal
M}d^2z~\tr\left[(\partial_zg)g^{-1}(\partial_{\bar
z}g)g^{-1}+2(\partial_zg)g^{-1}\omega_\bz\right] \nn \\
& &~~~~+\frac{k}{24 \pi} \int_{\cal M}\tr\left(g^{-1} dg\wedge g^{-1} dg\wedge
g^{-1} dg\right)
\label{2.9}\eea
Note that although the WZNW term in (\ref{2.9}) appears to be supported on the
bulk manifold $\cal M$, it actually turns out to be a total derivative and can
thus be written as a boundary term for the group elements $g(z,\bar z)\in
SL(2,\real)$.

The action (\ref{2.9}) is invariant under the left $SL(2,\real)$ gauge
transformations
\bea
g &\ra& h g \nn \\
e &\ra& h e h^{-1} \nn \\\beta&\to&h(d+\beta)h^{-1}\nn\\
\omega &\ra& h (d+ \omega ) h^{-1}\nn\\\Phi&\to&\Phi
\label{2.10}\eea
We shall consider here the perturbative phase of the theory in which there is a
non-zero condensate of the dreibeins, i.e. $\langle
e_\mu^a\rangle=\rho\,\delta_\mu^a$, where $\rho\neq0$ is a conformal factor. In
this phase the local $SL(2,\real)$ gauge symmetry (\ref{2.10}) is spontaneously
broken, and we can fix the pullback of the dreibein component $e_z$ to the
boundary to be
\be
e_z = \left( \begin{array}{cc}0&~\rho\\0&~0 \end{array} \right )
\label{2.13}\ee
If we consider another copy of $\partial\cal M$ of the opposite chirality and
orientation to that appearing above, and fix $e_\bz=e_z^\top$ on the opposite
boundary, then the total induced metric on the chirally-symmetric boundary is
$g_{z\bar{z}}=\tr( e_z e_{\bar{z}}) = \rho^2$. In the topological phase
$\langle e_\mu^a\rangle=0$, there is no background spacetime and again there
are no local graviton degrees of freedom. From the Cartan-Maurer equation
(\ref{spincon}) it follows that the fixed component of the spin-connection on
$\partial\cal M$ corresponding to the above choice of dreibein is
\be
\omega_{\bz} = \left(\begin{array}{cc}\partial_{\bz}\rho &~0 \\
0 & ~-\partial_{\bz}\rho \end{array} \right)
\label{omegafixed}\ee
However, we shall need to keep all fields arbitrary as yet until we carry out a
proper gauge-fixing of the path integral.

Note that the boundary term (\ref{2.5.1}), added to the conformally-coupled
topologically massive gravity action in (\ref{2.3.5}), involves only the
component $e_\bz$ of the dreibein field on $\partial\cal M$. This implies that,
despite the gauge choice (\ref{2.13}), there is still a residual abelian gauge
symmetry in (\ref{2.9}) defined by the action of $h$ in (\ref{2.10})
which restricted to the boundary lies in the Borel subgroup ${\cal B}_-$ of the
total $SL(2,\real)$ group consisting of lower triangular matrices,
\be
h = \left( \begin{array}{cc} 1 & ~0 \\
\xi &~1  \end{array} \right ).
\label{2.13.5}\ee
Let us now consider a local Gauss decomposition of the matrix-valued field $g$
in terms of the two Borel subgroups ${\cal B}_\pm$ and the Cartan subgroup of
$SL(2,\real)$,
\be
g = \left( \begin{array}{cc} 1 & ~0\\
\xi & ~1 \end{array} \right)
\left( \begin{array}{cc} \e^{-\varphi} &~ 0\\
0 &~ \e^{\varphi} \end{array} \right)
\left( \begin{array}{cc} 1 & ~\psi\\
0 &  ~1 \end{array} \right).
\label{2.11}
\ee
In terms of the variables in (\ref{2.11}) and the decomposition
\be
\omega_{z, \bz} = \left( \begin{array}{cc} \omega^3_{z,\bz} &
{}~\omega^+_{z,\bz} \\
\omega^-_{z,\bz} & ~-\omega^3_{z,\bz} \end{array} \right)
\label{2.14}\ee
of the spin-connection, the action (\ref{2.9}) can be evaluated using the
Polyakov-Wiegmann identity \cite{polwieg} for the WZNW action evaluated on the
product of the three matrix-valued fields in (\ref{2.11}). The resulting action
is invariant under the local residual Borel subgroup symmetry of the
holomorphic part of the dreibein condensate,
\bea
h&\to&h+\tau_-\theta\nn\\\omega_\bz&\to&\omega_\bz-\tau_-\,\partial_\bz\theta
\label{B+sym}\eea
where
\be
\tau_+=\left(\begin{array}{cc}0&~1\\0&~0\end{array}\right)
~~~~~~,~~~~~~\tau_-=\left(\begin{array}{cc}0&~0\\1&~0\end{array}\right)
~~~~~~,~~~~~~\tau_3=\left(\begin{array}{cc}1&~0\\0&~-1\end{array}\right)
\label{sigmadefs}\ee
are the generators of the fundamental representation of $SL(2,\real)$, and is
given by
\bea
S_T^-[g;e,\omega,\Phi,\beta]&=&S_{\rm
bulk}[e,\omega,\Phi,\beta]-2\kappa\oint_{\partial\cal
M}\Phi\,\partial_\perp\Phi+\frac{k}{4\pi}\oint_{\partial\cal M}d^2z~\partial_z
\varphi\,\partial_\bz\varphi\nn\\& &-\frac k{8\pi}\oint_{\partial\cal
M}d^2z~\left[2\omega_z^a\omega_{\bar
z}^a-4\omega^3_\bz\,\partial_z\varphi+\omega_{\bar
z}^-\left(\e^{-2\varphi}\,\partial_z\psi+\frac{8\pi\kappa}k\,\rho\,
\Phi^2\right)\right]
\label{2.123}\eea
In (\ref{2.123}) we have cancelled the boundary $\beta_ze_\bz$ term in
(\ref{2.7}) with the gauge dependence of the $\omega_{\bar z}e_z$ term and used
the extra abelian ${\cal B}_-$ gauge symmetry (\ref{B+sym}) to select the gauge
$\xi=0$. Having fixed the gauge, we also set the dreibein components to their
fixed boundary values (\ref{2.13}).

However, our construction thus far produces only one chiral sector of the
worldsheet theory on $\partial\cal M$. The full chirally-symmetric theory is
obtained by taking the geometry of the manifold $\cal M$ to be such that its
boundary is the connected sum $\partial{\cal M}=\partial{\cal
M}_+\,\#\,\partial{\cal M}_-$ of two isomorphic surfaces $\partial{\cal M}_\pm$
of opposite chirality and orientation (see fig. 1),\footnote{\baselineskip=12pt
For example, we could take ${\cal M}=\Sigma\times[0,1]$ where
$\Sigma\times\{0\}$ and $\Sigma\times\{1\}$ are surfaces of opposite chirality
and orientation.} and gauging the action with respect to both the lower Borel
subgroup ${\cal B}_-$ for the left-moving sector and the upper Borel subgroup
${\cal B}_+$ for the right-moving sector \cite{alex}. If, say, the action
(\ref{2.123}) is defined on $\partial{\cal M}_-$, then we can derive its
anti-holomorphic counterpart on $\partial{\cal M}_+$ in the same manner as
above by choosing the opposite chiral components in the boundary conditions
(\ref{2.5}), inducing an anti-chiral gauged WZNW action which is invariant
under right $SL(2,\real)$ gauge transformations, and thereby producing a
boundary action which is invariant under the residual ${\cal B}_+$ symmetry of
the anti-holomorphic part of the dreibein condensate. Taking into careful
account of the change of orientation on $\partial{\cal M}_+$ (fig. 1), this
leads to the anti-chiral version of (\ref{2.123}),
\bea
S_T^+[g;e,\omega,\Phi,\beta]&=&S_{\rm
bulk}[e,\omega,\Phi,\beta]+2\kappa\oint_{\partial\cal
M}\Phi\,\partial_\perp\Phi-\frac{k}{4\pi}\oint_{\partial\cal M}d^2z~\partial_z
\varphi\,\partial_\bz\varphi\nn\\& &+\frac k{8\pi}\oint_{\partial\cal
M}d^2z~\left[2\omega_z^a\omega_{\bar
z}^a-4\omega_z^3\,\partial_\bz\varphi+\omega_z^+\left(\e^{-2\varphi}\,
\partial_\bz\xi+\frac{8\pi\kappa}k\,\rho\,\Phi^2\right)\right]
\label{2.123bar}\eea
in the gauge $\psi=0$. Gluing the two contributions (\ref{2.123}) and
(\ref{2.123bar}) together \cite{alex}, we arrive in this way at the full
chirally symmetric action
\bea
S_T^{\rm sym}[g;e,\omega,\Phi,\beta]&=&S_{\rm
bulk}[e,\omega,\Phi,\beta]-2\kappa\oint_{\partial\cal
M}\Phi\,\partial_\perp\Phi\nn+\frac{k}{8 \pi}\oint_{\partial\cal
M}d^2z~\biggl[2\,\partial_z
\varphi\,\partial_\bz\varphi+\,\e^{-2\varphi}\,\omega_z^+\omega_\bz^-
\biggr.\nn\\& &\left.-\omega_z^-\omega_\bz^+-\frac{8\pi\kappa}k\,\rho\,
\Phi^2\left(\omega_z^++\omega_\bz^-\right)-2\left(\omega_z^3\omega_\bz^3-
2\omega_z^3\,\partial_\bz\varphi-2\omega_\bz^3\,\partial_z\varphi\right)\right]
\label{STsym}\eea
where we have used the remnant abelian ${\cal B}_+\times{\cal B}_-$ gauge
symmetry to fix the gauge $\xi=\psi=0$ on $\partial\cal M$. The appearence of
the $\e^{-2\varphi}\omega_z^+\omega_\bz^-$ term in (\ref{STsym}) comes from the
appropriate gluing required to produce the ${\cal B}_+\times{\cal B}_-$
symmetric WZNW model \cite{alex}. This equivalence follows from the fact
\cite{verlinde,carlip,kpz} that the boundary dynamics of ordinary topologically
massive gravity induce the full chirally symmetric Liouville theory.

The partition function of conformally-coupled topologically massive gravity is
defined by the gauge-fixed path integral
\bea
{\cal Z}&=&\int[de]~\Delta_{\rm FP}[e]~[d\beta]~\Delta_{\rm
FP}[\beta]~[d\omega]~\Delta_{\rm FP}[\omega]\nn\\&
&\times\int[d\Phi]~\int[dg]~\delta({\cal E}[\bar e])~\delta({\cal
L}[\,^g\beta])~\delta({\cal W}[\,^g\omega])~\e^{iS_{CTMG}[e,\omega,\Phi,\beta]}
\label{CTMGpartfn}\eea
where $^g\beta=g^{-1}\beta g+g^{-1}dg$ and $\bar e=g^{-1}eg$. Here $\Delta_{\rm
FP}$ denotes the Faddeev-Popov determinant, $\cal E$, $\cal L$ and $\cal W$ are
gauge-fixing functions, and $dg$ is the left-right invariant Haar measure on
$SL(2,\real)$. Following the steps which led to the effective action
(\ref{STsym}), we see that (\ref{CTMGpartfn}) can be written as
\bea
{\cal Z}&=&{\cal N}\int[d\bar e]~\delta({\cal E}[\bar e])~\Delta_{\rm FP}[\bar
e]~[d\bar\beta]~\delta({\cal L}[\bar\beta])~\Delta_{\rm
FP}[\bar\beta]~[d\bar\omega]~\delta({\cal W}[\bar\omega])~\Delta_{\rm
FP}[\bar\omega]~\int[d\Phi]~\e^{iS_{\rm bulk}[\bar
e,\bar\omega,\Phi,\bar\beta]}\nn\\&
&\times\int[d\varphi]~\det[\partial_z\,\partial_\bz]~\exp i\oint_{\partial\cal
M}\left[\frac
k{4\pi}\,d^2z~\partial_z\varphi\,\partial_\bz\varphi-2\kappa\,\Phi\,
\partial_\perp\Phi\right]\nn\\& &\times\int[d\omega_z^+]~[d\omega_\bz^-]~
\exp-\frac{ik}{8\pi}\oint_{\partial\cal M}d^2z~\left[\,\e^{-2\varphi}\,
\omega_z^+\omega_\bz^-+\frac{8\pi\kappa}k\,\rho\,\Phi^2\left(\omega_z^++
\omega_\bz^-\right)\right]\nn\\& &\times\int[d\omega_z^3]~[d\omega_\bz^3]~
\exp-\frac{ik}{4\pi}\oint_{\partial\cal M}d^2z~\left[\omega_z^3\omega_\bz^3-2
\omega_z^3\,\partial_\bz\varphi-2\omega_\bz^3\,\partial_z\varphi\right]
\label{partbulkbound}\eea
where the bars on the fields denote their bulk values which are parametrized by
their adjoint orbits under the $SL(2,\real)$ gauge group in (\ref{2.8}), and
the additional determinant comes from gauge-fixing the ${\cal B}_+\times{\cal
B}_-$ Borel symmetry. Here and in the following we will absorb irrelevant
(infinite) constants into the normalization factor $\cal N$. The functional
integration over the boundary spin-connection components in
(\ref{partbulkbound}) is Gaussian and yields a fluctuation determinant that can
be evaluated to give \cite{fluct}
\be
\prod_{(z,\bz)\in\partial\cal M}\e^{2\varphi(z,\bz)}={\cal
N}\exp\frac{i}{8\pi}\oint_{\partial\cal M}d^2z~Q\,\varphi\,R^{(2)}
\label{buscherfluct}\ee
where $R^{(2)}$ is the worldsheet scalar curvature of $\partial\cal M$. The
constant $Q$ in (\ref{buscherfluct}) is a regularization parameter which will
control the central charge of the induced Liouville theory. Upon rescaling
$\varphi\to\varphi/\sqrt{10k}$ and $Q\to\sqrt{10k}\,Q$ we arrive finally at
\bea
{\cal Z}&=&{\cal N}\int[d\bar e]~\delta({\cal E}[\bar e])~\Delta_{\rm FP}[\bar
e]~[d\bar\beta]~\delta({\cal L}[\bar\beta])~\Delta_{\rm
FP}[\bar\beta]~[d\bar\omega]~\delta({\cal W}[\bar\omega])~\Delta_{\rm
FP}[\bar\omega]\nn\\& &\times\int[d\Phi]~\e^{iS_{\rm bulk}[\bar
e,\bar\omega,\Phi,\bar\beta]}~\int[d\varphi]~\e^{iS_\partial[\varphi;\Phi]}
\label{partfinal}\eea
where
\be
S_\partial[\varphi;\Phi]=\oint_{\partial\cal M}d^2z~\left(\frac
1{8\pi}\,\partial_z\varphi\,\partial_\bz\varphi+\frac{Q}{8\pi}\,\varphi\,
R^{(2)}+\frac{2\pi\kappa^2}{k}\,\rho^2\,\Phi^4\,\e^{\sqrt{2/5k}\,\varphi}
\right)-2\kappa\oint_{\partial\cal M}\Phi\,\partial_\perp\Phi
\label{defliouville}\ee
Note that the dreibein condensate parameter $\rho^2=\sqrt{g}$ cancels in the
first two terms of (\ref{defliouville}) because of the additional contractions
with the induced metric $g^{z\bz}$ required for worldsheet general covariance.

The boundary induced action (\ref{defliouville}) is very similar to that for
two-dimensional quantum gravity, with the field $\varphi$ which parametrizes
the Cartan subgroup of the three-dimensional Lorentz symmetry group
$SL(2,\real)$ identified as the Liouville field. However, there are two crucial
differences. The first one is that although the three-dimensional scalar field
$\Phi$ is constant on $\partial\cal M$, it is scale-dependent, and it therefore
defines a dynamical conformal deformation of the cosmological constant operator
in (\ref{defliouville}). The vacuum expectation value of the scalar field
$\Phi$ determines the two-dimensional cosmological constant $\mu$, i.e. the
scale of Liouville theory, which with the normalization in (\ref{defliouville})
is given by
\be
\mu=\mbox{$\frac1{10}$}\,M_g^2
\label{2dcosm}\ee
where $M_g$ is the topological graviton mass (\ref{plmass}). Thus the natural
scale of the two-dimensional boundary theory is determined by that of the bulk
three-dimensional theory \cite{kogan}, so that the critical value $\mu=0$ comes
from unbroken $SO(2,1)$ conformal symmetry of the three-dimensional dynamics.
The second difference is that the additional boundary term in
(\ref{defliouville}) which is independent of the Liouville field depends on the
bulk value of $\Phi$ in a neighbourhood of the boundary $\partial\cal M$. This
term does not affect the Weyl transformation properties of the induced
two-dimensional theory and merely serves to maintain the conformal symmetry of
the bulk part of the three-dimensional theory (\ref{partfinal}). In particular,
it prevents the complete factorization of bulk and boundary degrees of freedom
in the quantum field theory (in contrast to the usual cases
\cite{carlip,witten}), so that the theory (\ref{defliouville}) is intrinsically
three-dimensional in origin. This feature is important to remember when
analysing certain aspects such as the conformal invariance properties of the
boundary theory. We shall refer to the model defined by the action
(\ref{defliouville}) as ``deformed Liouville theory''.

\section{Induced Dilaton Field and Conformal Symmetry}

In the previous section we have shown that, in the phase with spontaneous
breaking of the conformal and Lorentz symmetries of the three-dimensional
quantum field theory, conformally-coupled topologically massive gravity induces
a deformed Liouville theory (\ref{defliouville}) on the boundary of the
three-dimensional spacetime. The dynamical scaling constant
$\Phi|_{\partial\cal M}$ can be removed from the cosmological constant term in
(\ref{defliouville}) by the shift $\varphi\to\varphi-\frac1{\alpha_+}\ln\Phi^4$
of the Liouville field, where $\alpha_+=\sqrt{\frac2{5k}}$, so that the action
(\ref{defliouville}) can be written as
\bea
S_\partial[\varphi;\Phi]&=&\oint_{\partial\cal M}d^2z~\left(\frac
1{8\pi}\,\partial_z\varphi\,\partial_\bz\varphi+\frac{Q}{8\pi}\,\varphi\,
R^{(2)}+\frac{\rho^2\,\hat\mu}{8\pi\alpha_+^2}\,\e^{\alpha_+\varphi}\right)-
\frac Q{\alpha_+}\,S_D[\Phi]-2\kappa\oint_{\partial\cal M}\Phi\,
\partial_\perp\Phi\nn\\
\label{defliouvilleshift}\eea
where $\hat\mu=\mu/\langle\Phi^2\rangle^2$ is a fiducial worldsheet scale and
\be
S_D[\Phi]=\frac1{4\pi}\oint_{\partial\cal
M}d^2z~\left(\ln\Phi^2\right)\,R^{(2)}
\label{dilatonaction}\ee
is the usual form of the action for the dilaton field in a string sigma-model.
In (\ref{defliouvilleshift}) we have incorporated an arbitrary constant
$\alpha_+$ in the cosmological constant operator so that it may be made
marginal (i.e. $\alpha_+$ is adjusted so that $\e^{\alpha_+\varphi}$ has
conformal dimension 1) \cite{kogan}. This term is then to be thought of as
being induced by higher-loop effects, in analogy to the situation in
two-dimensional quantum gravity \cite{ddk} whereby quantum fluctuations also
change the factor parametrizing the cosmological constant operator. Since
$\Phi$ can vary in the directions normal to $\partial\cal M$ in $\cal M$, its
value in the bulk parametrizes the worldsheet dilaton field $\ln\Phi^2$. The
fact that it is constant here owes to the property that the present model
induces a theory of pure two-dimensional quantum gravity, i.e. it lives in a
zero-dimensional target space. In the next section we will consider the
coupling of the theory (\ref{defliouvilleshift}) to dynamical matter fields
from the bulk point of view and hence the properties of the theory when
embedded into higher-dimensional spacetimes, more precisely in flat toroidal
backgrounds where again the dilaton field should be constant. In this section
we shall use conformal invariance to determine a relation between the
parameters of the Liouville action in (\ref{defliouvilleshift}) and the induced
dilaton field.

For this, we parametrize the field $\Phi$ as
\be
\Phi(x)=\e^{\gamma\phi(x)}~~~~~~,~~~~~~x\in\cal M
\label{phidef}\ee
In the functional form (\ref{phidef}) the dilaton field is thus related to the
external tachyon operator in target space whose vacuum expectation value is the
scale $\hat\mu$ of the induced two-dimensional theory. Thus the present
framework gives a relationship between the bulk dilaton field and the target
space tachyon operator, such that the background tachyon field is responsible
for the spontaneous breaking of the three-dimensional conformal symmetry. The
constant $\gamma$ will be fixed by demanding that the two-dimensional action
(\ref{defliouvilleshift}) be conformally-invariant, as it should be due to the
Weyl invariance of the bulk theory.

Under a shift of the linear dilaton field,
\be
\phi\to\phi+\delta\phi~~~~~~{\rm with}~~\delta\phi\Bigm|_{\partial\cal M}={\rm
constant}
\label{phishift}\ee
corresponding to a global scale transformation of the worldsheet, the action
(\ref{defliouvilleshift}) changes as
\be
S_\partial[\varphi;\Phi]\to S_\partial[\varphi;\Phi]-\frac{2\gamma
Q}{\alpha_+}\,\chi_E(\partial{\cal
M})\,\delta\phi+2\kappa\left(1-\e^{2\gamma\delta\phi}\right)\oint_{\partial\cal
M}\Phi\,\partial_\perp\Phi-2\kappa\gamma\,\e^{2\gamma\delta\phi}
\oint_{\partial\cal M}\Phi^2\,\partial_\perp\delta\phi
\label{actionshift}\ee
where $\chi_E(\partial{\cal M})=\frac1{4\pi}\oint_{\partial\cal
M}d^2z\,R^{(2)}=2(1-h_{\partial\cal M})$ is the Euler character of the Riemann
surface $\partial\cal M$ and $h_{\partial\cal M}$ is its genus. When the theory
(\ref{defliouvilleshift}) is coupled to a string sigma-model (as we will do in
the next section), the dilatonic shift should be absorbed into a redefinition
of the string coupling constant as $g_s\to\e^{\delta\phi}\,g_s$. Since the
latter quantity appears in the string perturbation expansion as
$g_s^{2(1-h_{\partial\cal M})}$, this means that, if we assume that $\Phi$ is
constant everywhere on $M$, then the linear variation in $\phi$ should be
$2(1-h_{\partial\cal M})\delta\phi$ which leads to the constraint
\be
-\frac{2\gamma Q}{\alpha_+}=1
\label{parconstr}\ee
on the parameters of the theory (\ref{defliouvilleshift}). However, in
(\ref{actionshift}) there is generally an extra conformally non-invariant term
which is a remnant from the embedding of the worldsheet $\partial\cal M$ into
the bulk of the three-manifold $\cal M$. For the time being we shall ignore
this extra three-dimensional piece to illustrate how one can fix the
parameters. When Liouville gravity is coupled to conformal matter fields of
central charge $c\leq1$, the requirement that the total quantum action be
conformally invariant fixes the parameters of the Liouville action as
\cite{ddk}
\be
Q=\sqrt{\frac{25-c}3}~~~~~~,~~~~~~\alpha_+=-\frac Q2+\sqrt{\frac{1-c}{12}}
\label{confpar}\ee
This shows that the field $\phi(x)$ in (\ref{phidef}) can be identified as a
string dilaton provided we take
\be
\gamma=\frac14\left(1-\sqrt{\frac{1-c}{25-c}}\,\right)
\label{gammafix}\ee

\section{Sigma-model Couplings and Quantum Duality Transformations}

We will now consider the coupling of deformed Liouville gravity to conformal
matter fields. From the point of view of a topological membrane, this means
that we add propagating gauge field degrees of freedom to the bulk, in addition
to the gravity and scalar fields. As always this is done in a conformally
invariant way. The simplest situation is described by the bulk action
\be
{\cal S}[e,\omega,\Phi,A]=S_{CTMG}[e,\omega;\Phi]+S_{CTMGT}^{[U(1)]}[A;e,\Phi]
\label{bulkgauge}\ee
where
\be
S_{CTMGT}^{[U(1)]}[A;e,\Phi]=-\frac{1}{4}\int_{\cal
M}d^3x~\left(\sqrt{g}~g^{\mu\rho} g^{\nu\sigma}\,\frac{1}{\Phi^2}\,F_{\mu
\nu}F_{\rho \sigma}-\frac{2m}\pi\,\epsilon^{\mu \nu \lambda}
A_\mu\,\partial_\nu A_\lambda \right)
\label{1.2}\ee
is the action for $U(1)$ topologically massive gauge theory conformally-coupled
to the scalar field $\Phi$. Here $A$ is an abelian gauge connection on $\cal M$
and $F=dA$ is its field strength. The non-polynomial coupling to the scalar
field $\Phi$ ensures the invariance of the action (\ref{1.2}) under the Weyl
transformations (\ref{confmetric}). When $\langle\Phi^2\rangle=0$ the model
(\ref{1.2}) is equivalent to pure Maxwell theory which has a massless photon.
However, when the conformal symmetry is spontaneously broken there is a massive
propagating photon in the bulk with topological mass
\be
M_p=\frac{m\,\langle\Phi^2\rangle}{\pi}
\label{massgap1}\ee

When $\partial{\cal M}\neq\emptyset$, the only gauge non-invariant term in
(\ref{1.2}) is the Chern-Simons action and therefore we expect the same WZNW
model to be induced on the boundary as in the case of a non-conformal bulk
coupling. Indeed, because of the Dirichlet boundary condition (\ref{bc1}), the
variation of the action (\ref{1.2}) restricted to the boundary is
\be
\delta S_{CTMGT}^{[U(1)]}[A;e,\Phi]\Bigm|_{\partial\cal M}=\oint_{\partial\cal
M}\left(\Pi^z\delta A_z+\Pi^\bz\delta A_\bz\right)
\label{CTMGTbdryvar}\ee
where
\be
\Pi^{z,\bz}=\frac1{\Phi^2}\,{\sqrt g}\,F^{z\perp}+\frac{2m}\pi\,\sqrt
g\,g^{z\bz}A_{\bz,z}
\label{canmom}\ee
is the canonical momentum conjugate to the gauge field. This is the same
boundary variation that occurs in the usual case \cite{carlip,cooper} and one
can therefore proceed to factorize the gauge-fixed path integral for the gauge
theory (\ref{1.2}) into bulk and boundary components. When the $U(1)$ gauge
group is compact, the bulk theory (\ref{bulkgauge}) induces the two-dimensional
boundary action
\be
{\cal
S}_\partial[\varphi,\phi,\theta]=S_\partial[\varphi;\phi]+
S_{XY}^{[S^1]}[\theta]
\label{totalctiongauge}\ee
where
\be
S_{XY}^{[S^1]}[\theta]=\frac m{2\pi}\oint_{\partial\cal
M}d^2z~\partial_z\theta\,\partial_\bz\theta
\label{XYaction1}\ee
is the linear sigma-model action with $\theta\in S^1$ the pure gauge part of
$A$ on the boundary $\partial\cal M$. The gauge field Chern-Simons coefficient
is related to the radius $R$ of the circle $S^1$ by
\be
m=\frac{R^2}{\alpha'}
\label{mradius}\ee
Thus the action (\ref{bulkgauge}) induces a coupling of the deformed Liouville
theory (\ref{defliouvilleshift}) to the $c=1$ conformal field theory of the
$XY$ model.

Let us now consider the behaviour of the theory (\ref{bulkgauge}) under a
$T$-duality transformation $R\to\alpha'/R$ of the target space $S^1$ of the
$XY$ model (\ref{XYaction1}), which is a symmetry of the two-dimensional
quantum field theory. This mapping is equivalent to the transformation
\be
m\to m^*=\frac1m
\label{Tdualm}\ee
of the Chern-Simons coefficient of the bulk theory (\ref{bulkgauge}). The
target space $T$-duality transformation therefore arises from an $S$-duality
transformation of the three-dimensional quantum field theory. The effects of
this transformation in topologically massive gauge theory have been extensively
studied in \cite{tmgtdual,annals}. Although it is not a precise symmetry of the
bulk theory,\footnote{\baselineskip=12pt Under the $T$-duality transformation
(\ref{Tdualm}), the topologically massive gauge theory action (\ref{1.2}) is
mapped into a Chern-Simons-Proca gauge theory with the same mass
(\ref{massgap1}). Moreover, the mapping interchanges winding numbers of matter
fields and monopole numbers in the spectrum of the quantum gauge theory. See
\cite{annals} for details.} in the same sense as the way it acts on the
boundary theory, it does provide a one-to-one mapping between the spectrum of
the quantum gauge theory and its dual, i.e. it preserves the Landau level
structure of states. Since in the present case the mass gap (\ref{massgap1})
between states involves the dynamical scalar field $\Phi$, we can demand that
it be invariant under the mapping (\ref{Tdualm}). More precisely, we introduce
a dual scalar field $\Phi^*(x)$ on $\cal M$ so that
\be
M_p=\frac{m\,\langle\Phi^2\rangle}\pi=\frac{m^*\langle\Phi^{*\,2}\rangle}\pi
\label{Mpinv}\ee
This leads to the $T$-duality transformation law
\be
\Phi\to\Phi^*=m\,\Phi
\label{Phidual}\ee
or, using (\ref{phidef}) and (\ref{gammafix}), the transformation of the linear
dilaton
\be
\phi\to\phi^*=\phi+4\ln\frac{R^2}{\alpha'}
\label{phidual}\ee

Eq. (\ref{phidual}) differs by a factor of $\frac18$ from the usual
transformation law (\ref{buscherlaw}) for the dilaton field under $T$-duality.
The discrepency can be traced back to the extra local operator that appears in
the deformed Liouville action (\ref{defliouvilleshift}) which provides a
coupling to the bulk degrees of freedom. This means that the scale-invariance
arguments which we used to fix the constant $\gamma$ in (\ref{gammafix}) are
not precisely valid in the present case (c.f. (\ref{actionshift})). As we have
stressed in sections 2 and 3, the action (\ref{defliouvilleshift}) is
intrinsically three-dimensional in origin, so that the usual arguments of
two-dimensional quantum gravity require some modification that would presumably
change the numerical value of $\gamma$ given in section 3 and give agreement
with the standard results. In any case, we take (\ref{phidual}) to be the
three-dimensional version of the dilaton transformation law.

In fact, the above arguments lead immediately to an intimate relationship
between $T$-duality and $S$-duality. On the worldsheet boundary the $T$-duality
and $S$-duality mappings are given, respectively, by
\bea
T&:&~R\to\frac{\alpha'}R~~~~,~~~~g_s^2\to({\rm const.})\cdot
g_s^2\nn\\S&:&~g_s\to\frac1{g_s}
\label{bdrydual}\eea
where $g_s$ is the string coupling constant (\ref{gsdef}). On the other hand,
in order to keep the mass gap of the bulk three-dimensional theory invariant
under the mapping (\ref{Tdualm}), the dilaton condensate must change (up to a
constant) like $\langle\Phi^2\rangle\to1/\langle\Phi^2\rangle$. Thus in the
bulk the analogs of the $S$-duality and $T$-duality transformations are
respectively
\bea
\bbox{S}&:&~m\to\frac1m~~~~,~~~~\Phi\to
m\,\Phi\nn\\\bbox{T}&:&~\Phi^2\to\frac{\rm const.}{\Phi^2}
\label{bulkdual}\eea
Although in the boundary theory the $S$-duality and $T$-duality symmetries
appear to be unrelated, the bulk theory unifies them via the change of role of
the scalar field $\Phi$ in the bulk-boundary correspondence, i.e. $\bbox{S}\sim
T$ and $\bbox{T}\sim S$. Specifically, the duality transformations
(\ref{bulkdual}) both change the topological masses of the three-dimensional
theory (photon mass for $\bbox{S}$ and graviton mass for $\bbox{T}$). This
gives a remarkable dynamical equivalence between the quantum geometry of the
target space and the non-perturbative properties of the quantum string theory.

Within the present three-dimensional framework there is an interesting
generalization of this construction for higher-dimensional toroidal string
compactifications. For this, we consider a compact $U(1)^D$ conformally-coupled
topologically massive gauge theory with potentials $A^I$ and associated field
strengths $F^I=dA^I$, $I=1,\dots,D$. The action is
\be
S_{CTMGT}^{[U(1)^D]}[A;e,\Phi]=-\frac{1}{4}\sum_{I,J}\int_{\cal
M}d^3x~\left(\sqrt{g}~g^{\mu\rho} g^{\nu\sigma}\,\frac{1}{\Phi^2_{IJ}}\,F_{\mu
\nu}^IF_{\rho \sigma}^J-\frac{2}\pi\,K_{IJ}\,\epsilon^{\mu \nu \lambda}
A_\mu^I\,\partial_\nu A_\lambda^J \right)
\label{1.2D}\ee
where $K_{IJ}$ is a non-degenerate constant $D\times D$ matrix, and the
functions $\Phi_{IJ}(x)=\Phi_{JI}(x)$ each transform under restricted
three-dimensional conformal transformations according to (\ref{confmetric}).
The action (\ref{1.2D}) induces the $c=D$ conformal field theory of the $XY$
model
\be
S_{XY}^{[T^D]}[\theta]=\frac1{2\pi}\sum_{I,J}\oint_{\partial\cal
M}d^2z~K_{IJ}\,\partial_z\theta^I\,\partial_\bz\theta^J
\label{XYactionD}\ee
where the pure gauge degrees of freedom $\theta^I$ of the gauge fields $A^I$
live in a $D$-torus $T^D$. This identifies the Chern-Simons coefficient matrix
as
\be
K_{IJ}=\frac1{\alpha'}\Bigl(G_{IJ}+B_{IJ}\Bigr)
\label{KGB}\ee
where $G_{IJ}$ and $B_{IJ}$ are the target space graviton and antisymmetric
tensor condensates, respectively. In the phase of non-zero meson condensates,
it follows from the gauge field equations of motion that the local propagating
degrees of freedom of the model (\ref{1.2D}) can be characterized by the mass
matrix
\be
M_{IJ}=\frac1{\pi\alpha'}\sum_LG_{JL}\left\langle\Phi_{IL}^2\right\rangle
\label{massgapD}\ee
Here we have used the fact that the bulk part of the Chern-Simons action in
(\ref{1.2D}), and hence the gauge field propagator, depends only on the
symmetric part of the matrix $K_{IJ}$, i.e. on the metric tensor $G_{IJ}$ of
$T^D$. This can be seen via an integration by parts of the Chern-Simons
three-form.

A $T$-duality transformation of the quantum field theory (\ref{XYactionD})
corresponds to inversion of the Chern-Simons coefficient matrix
$K_{IJ}\to(K^{-1})_{IJ}$, or in terms of the metric of $T^D$
\be
G_{IJ}\to G^*_{IJ}=\left[(K^\top)^{-1}\,G\,K^{-1}\right]_{IJ}
\label{TdualG}\ee
Defining a duality transformation $\Phi_{IJ}\to\Phi_{IJ}^*$ such that the mass
matrix (\ref{massgapD}) is preserved by (\ref{TdualG}), we find
\be
\Phi_{IJ}^2\to\Phi^{*\,2}_{IJ}=\sum_L\left[K\,G^{-1}\,K^\top\,G\right]_{JL}
\Phi_{IL}^2
\label{PhidualD}\ee
If we now define
\be
\Phi(x)=\det_{I,J}\left[\Phi_{IJ}(x)\right]=\e^{\phi(x)/4}
\label{dilatondefD}\ee
then (\ref{PhidualD}) implies that the linear dilaton field $\phi$ transforms
under toroidal $T$-duality transformations as
\be
\phi\to\phi^*=\phi+2\ln\frac{\det_{I,J}\left[G_{IJ}\right]}{\det_{I,J}
\left[G^*_{IJ}\right]}
\label{phidualD}\ee
Modulo the usual factor of $\frac18$, (\ref{phidualD}) has the precise form of
the dilaton transformation law for generic toroidal compactifications
\cite{phimulti}.

Eq. (\ref{PhidualD}) defines a dilaton transformation law that has no analog in
the induced two-dimensional theory and is purely three-dimensional in origin.
It comes from the $D>1$ local gauge field excitations characterized by
(\ref{massgapD}). One can think of this generalization as providing an
independent dilaton field $\phi_I(x)$ along each direction of the target space,
where $\e^{\phi_I(x)/4}$ are the eigenvalues of the symmetric matrix function
$\Phi_{IJ}(x)$, such that $\phi_I(x)$ controls the size of the compactified
direction $I$. The canonical dilaton in (\ref{dilatondefD}) is then the average
of these fields over the various directions and it mediates the scaling
properties of the entire spacetime as a whole. It is a highly non-trivial
property of the three-dimensional models that the local dynamics of the bulk
theory, measured here by (\ref{massgap1}) and (\ref{massgapD}), conspire to
yield the anticipated boundary properties of the induced two-dimensional sigma
models. For instance, for the transformation law (\ref{phidualD}) to hold it is
crucial that {\it only} the symmetric part of the Chern-Simons coefficient
matrix $K_{IJ}$ appear in (\ref{massgapD}). The present approach thus yields a
natural dynamical and geometrical origin for the dilaton in string
sigma-models.

\section{Dynamical String-Coupling Generation}

As we have discussed, the dilaton field of deformed Liouville theory couples to
the bulk of the three-manifold $\cal M$ and is strictly speaking not a constant
field in target space. The bulk dynamics of conformally-coupled topologically
massive gravity control the worldsheet properties of this field, and this
raises the possibility that the scalar field $\Phi$ may in fact generate a
dynamical string coupling constant (\ref{gsdef}). The most elegant way of
exploring this possibility is to consider the induced two-dimensional theory
that arises when, instead of the Dirichlet boundary condition (\ref{bc1}), one
imposes a Neumann boundary condition on the scalar field $\Phi(x)$ on
$\partial\cal M$,
\be
\partial_\perp\Phi=0
\label{neumannbc}\ee
In other words, the field $\Phi$ is constant in a neighbourhood of
$\partial\cal M$ in $\cal M$ but it is an arbitrary function $\Phi(z,\bz)$ on
the worldsheet $\partial\cal M$.\footnote{\baselineskip=12pt A more
mathematical way of saying this is that $\Phi$ defines a global section of the
normal bundle over $\partial\cal M$ in $\cal M$.} Now the conformal symmetry
(\ref{confmetric}) in the bulk is broken down to the subgroup of local
$SO(2,1)$ conformal transformations which are constant along the directions
normal to $\partial\cal M$ in $\cal M$. Such a restriction on the local
conformal symmetry group of the three-dimensional theory is very natural from
the point of view of the fact that it should induce the local scale invariance
of the two-dimensional boundary theory. Moreover, the additional boundary term
in (\ref{boundphi}) vanishes with the choice of Neumann boundary condition.

However, there are now extra boundary terms associated with the non-constant
field $\Phi|_{\partial\cal M}$ coming from the additional boundary terms in
(\ref{2.7}) and (\ref{2.9}) due to the non-vanishing variation $\delta\Phi$ on
$\partial\cal M$. The simplest way to incorporate such terms is to consider a
boundary Weyl transformation of the action (\ref{2.7}). From (\ref{spintrans})
and (\ref{2.13}) it follows that the boundary components of the spin-connection
transform as
\be
\omega_z\to\omega_z+\mbox{$\frac\gamma2$}\,\tau_3\,\partial_\bz\phi
~~~~~~,~~~~~~\omega_\bz\to\omega_\bz-\mbox{$\frac\gamma2$}\,\tau_3\,
\partial_z\phi
\label{bdryspintransf}\ee
It follows that the effect of allowing the field $\phi$ to vary along the
boundary is to shift the diagonal components $\omega_{z,\bar z}^3$ of the
spin-connection. The action (\ref{2.7}) is therefore modified by the shift
(\ref{bdryspintransf}) to
\bea
S_T[e,\omega,\phi,\beta]&=&S_{\rm
bulk}[e,\omega,\phi,\beta]+\oint_{\partial\cal
M}d^2z~\biggl\{\e^{2\gamma\phi}~\tr(e_\bz\beta_z-\kappa
e_z\omega_\bz)\biggr.\nn\\& &\left.-\frac
k{8\pi}\left[\omega_z^+\omega_\bz^-+\omega_\bz^+\omega_z^-+2
\left(\omega_z^3+\mbox{$\frac\gamma2$}\,\partial_\bz\phi\right)
\left(\omega_\bz^3-\mbox{$\frac\gamma2$}\,\partial_z\phi\right)\right]\right\}
\label{STmod}\eea
The additional contributions involving the dilaton field in (\ref{STmod}) are
most transparent when written in terms of the worldsheet $T$-dual field
$\widetilde{\phi}$ defined by
\be
\partial_z\phi=\partial_\bz\widetilde{\phi}
~~~~~~,~~~~~~\partial_\bz\phi=-\partial_z\widetilde{\phi}
\label{dualphidef}\ee
The dual dilaton field $\widetilde{\phi}$ is only locally defined on the
boundary $\partial\cal M$ according to the Poincar\'e lemma. Substituting
(\ref{dualphidef}) into (\ref{STmod}), reflecting the anti-holomorphic diagonal
component $\omega_\bz^3\to-\omega^3_\bz$, and integrating the
$\omega^3\,d\widetilde{\phi}$ cross-term by parts (ignoring the singularities
in the definition of $\widetilde{\phi}$), we arrive at
\bea
S_T[e,\omega,\phi,\beta]&=&S_{\rm
bulk}[e,\omega,\phi,\beta]+\oint_{\partial\cal
M}d^2z~\biggl\{\e^{2\gamma\phi}~\tr(e_\bz\beta_z-\kappa
e_z\omega_\bz)\biggr.\nn\\& &\left.-\frac
k{8\pi}\left[\omega_z^+\omega_\bz^-+\omega_\bz^+\omega_z^-
-2\omega_z^3\omega_\bz^3-\gamma\,\widetilde{\phi}\,R_{z\bz}+
\mbox{$\frac{\gamma^2}2$}\,\partial_z\widetilde{\phi}\,\partial_\bz
\widetilde{\phi}\right]\right\}
\label{STdualphi}\eea
where $R_{z\bz}=\partial_z\omega_\bz^3-\partial_\bz\omega_z^3$. The advantage
of this dual representation is that the scalar field $\widetilde{\phi}$ couples
in a conformally-invariant way to the induced worldsheet geometry.

The shift (\ref{bdryspintransf}) also produces a coupling of $\widetilde{\phi}$
to the Liouville field which comes from the gauged WZNW action (\ref{2.9}) (see
(\ref{partbulkbound})). After the rescaling $\varphi\to\varphi/\sqrt{10k}$, we
find that the boundary action (\ref{defliouville}) acquires the additional term
\be
\frac\gamma{20\pi\alpha_+}\oint_{\partial\cal
M}d^2z~\left(-\frac\gamma{2\alpha_+}\,\partial_z\widetilde{\phi}\,
\partial_\bz\widetilde{\phi}+\frac1{\alpha_+}\,\widetilde{\phi}\,R^{(2)}+
\partial_z\widetilde{\phi}\,\partial_\bz\varphi+\partial_\bz
\widetilde{\phi}\,\partial_z\varphi\right)
\label{phidualadd}\ee
where we have defined $k=2/5\alpha_+^2$. Furthermore, when one shifts the
Liouville field $\varphi$ to write the cosmological constant operator of the
Liouville part of the total action in standard form, there are additional
kinetic terms induced for the dilaton field $\phi$. The total boundary action
in this case is thus
\bea
S_\partial[\varphi;\phi,\widetilde{\phi}]&=&\oint_{\partial\cal
M}d^2z~\left(\frac 1{8\pi}\,\partial_z\varphi\,
\partial_\bz\varphi+\frac{Q}{8\pi}\,\varphi\,R^{(2)}+
\frac{\rho^2\,\hat\mu}{8\pi\alpha_+^2}\,\e^{\alpha_+\varphi}\right)-
\frac{2\gamma Q}{\alpha_+}\,S_D[\phi]+\frac\gamma{5\alpha_+^2}\,
S_D[\widetilde{\phi}]\nn\\& &+\oint_{\partial\cal
M}d^2z~\left[\frac{2\gamma^2}{\pi\alpha_+^2}\,\partial_z\phi\,
\partial_\bz\phi-\frac\gamma{2\pi\alpha_+}\Bigl(\partial_z\phi\,
\partial_\bz\varphi+\partial_\bz\phi\,\partial_z\varphi\Bigr)
\right]\nn\\& &+\frac\gamma{20\pi\alpha_+}\oint_{\partial\cal M}d^2z~
\left(-\frac\gamma{2\alpha_+}\,\partial_z\widetilde{\phi}\,\partial_\bz
\widetilde{\phi}+\partial_z\widetilde{\phi}\,\partial_\bz\varphi+
\partial_\bz\widetilde{\phi}\,\partial_z\varphi\right)\nn\\& &-
\frac{\gamma^2}{5\pi\alpha_+^2}\oint_{\partial\cal M}d^2z~
\left(\partial_z\phi\,\partial_\bz\widetilde{\phi}+\partial_\bz\phi\,
\partial_z\widetilde{\phi}\right)
\label{actiongsdyn}\eea
where $S_D[\phi]=\frac1{4\pi}\oint_{\partial\cal M}d^2z~\phi\,R^{(2)}$ is the
usual dilaton action.

In the resulting path integral the bulk and boundary degrees of freedom are now
completely decoupled and the action (\ref{actiongsdyn}) defines a purely
two-dimensional field theory (in contrast to the previous case). We may think
of this action as defining the local dynamics of a string-coupling field
$g_s(z,\bz)$. This field couples to the two-dimensional worldsheet quantum
gravity and its action (\ref{actiongsdyn}) involves the corresponding
(singular) worldsheet dual field. In fact, the action (\ref{actiongsdyn}) is
invariant under the worldsheet duality transformation
\be
\phi\to-\mbox{$\frac{\alpha_+}{10Q}$}\,\widetilde{\phi}
\label{wsTdualmap}\ee
provided that the parameters of the Liouville part of the total action are
fixed as
\be
Q=\alpha_+=-\mbox{$\frac25$}
\label{Tdualfixed}\ee
Thus in the phase (\ref{Tdualfixed}) of unbroken worldsheet $T$-duality
symmetry, the action (\ref{actiongsdyn}) describes a new form of non-unitary
matter fields coupled to two-dimensional quantum gravity. It would be
interesting to explore if this phase is related to the strong-coupling phase of
Liouville theory whose properties are largely unknown.

It is in this way that the bulk dynamics of topologically massive gravity can
induce a dynamical scale parameter which, when coupled to string sigma-models
as described in section 4, allows one to dynamically control the size of the
radii of the compactified dimensions of the target space. Via its coupling to
$\varphi$ in (\ref{actiongsdyn}), it also controls the extra ``time'' direction
induced by the Liouville field. The creation of the field $\Phi(z,\bz)$ thereby
give an induced theory which is more general than string theory, i.e. it puts
the topological membrane approach into a more unified setting like $M$-theory
\cite{wittenM} or other extensions of string theory. In particular, since the
theory (\ref{actiongsdyn}) involves both the field $\phi$ and its (independent)
dual $\widetilde{\phi}$, the induced worldsheet model appears to yield two
extra dimensions in target space and may play a role in understanding the extra
dimensionality of $F$-theory \cite{Ftheory}. Moreover, if we interpret the
appearence of both $\phi$ and $\widetilde{\phi}$ as implying the existence of
two independent Liouville fields, and hence two ``times'', then the
construction of this section can be thought of as giving a worldsheet origin,
via topological membranes, for models with two time evolution parameters
\cite{bars} which have symmetry groups that coincide with those of $AdS_D$
spacetimes. The model (\ref{actiongsdyn}) therefore also suggests a natural
dynamical and geometrical origin, in terms of topological membranes, for the
believed correspondence between conformal field theories and supergravity on
anti-de Sitter spacetimes \cite{adscft}. It would be interesting to exploit
properties of the worldsheet theory (\ref{actiongsdyn}), such as its worldsheet
$T$-duality symmetry, to explore features of the spacetime in connection with
$M$-theory and these other generalizations. In fact, the present construction
can be thought of as giving a dynamical origin to the appearence of extra
dimensions in this framework and as illustrating how the dynamics of Liouville
gravity appear in the 11-dimensional model. In this framework these dynamical
components are all described by basic string degrees of freedom which are
induced by the bulk dynamics of the topological membrane, thereby illustrating
the relevance of both string dynamics and topological membranes to the full
dynamics of $M$-theory. It would be most interesting to see what this implies
for a target space Lagrangian formalism for the latter theory.

\begin{figure}[htb]
\baselineskip=12pt
\vspace{30pt}
\epsfxsize=3in
\epsfysize=2in
\centerline{\epsfbox{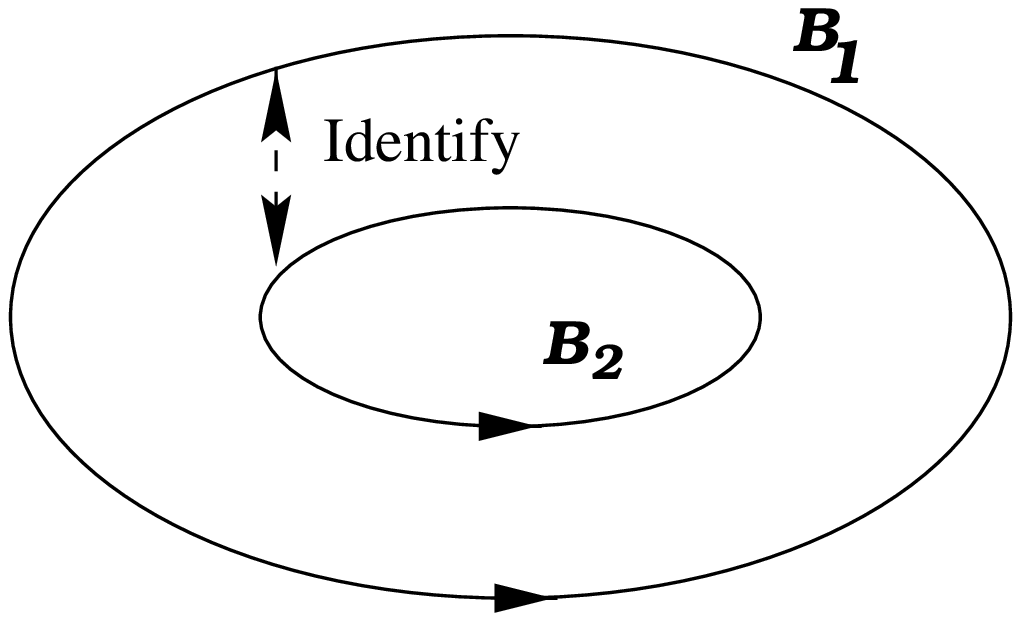}}
\vspace{30pt}
\caption{The annular spatial slice, with two boundaries $B_1=\partial{\cal
M}_+$ and $B_2=\partial{\cal M}_-$, corresponding to the three-geometry ${\cal
M}=\Sigma\times\real$ where $\Sigma$ is an annulus. Note that the boundaries
are oriented oppositely so that they can be glued together.}
\vspace{30pt}
\label{fig1}\end{figure}

\end{document}